\documentclass[12pt]{article}

\usepackage{epsfig}
\usepackage{amsmath}
\usepackage{soul,color}
\usepackage{bm}
\usepackage{epsfig}
\usepackage{natbib}
\usepackage{latexsym, graphicx, alltt}

\renewenvironment{output}
  {\small \def\baselinestretch{1.0}
   \begin{alltt}}
  {\end{alltt} \normalsize}
\newcommand{\tr}{^{\prime}}
\def\b#1{\mbox{\boldmath $#1$}}    
\def\bl#1{\mbox{\footnotesize\boldmath {$#1$}}} 

\def\cg#1{\mbox{${\cal #1}$}}      
\def\cgl#1{\mbox{\scriptsize {${\cal #1}$}}}

\newcommand{\ot}{\mbox{$\:\otimes \:$}}
\newcommand{\diag}{{\rm diag}}      
\renewcommand{\th}{\theta}
\newcommand{\Th}{\Theta}

\newcommand{\be}{\beta}
\newcommand{\de}{\delta}
\newcommand{\la}{\lambda}

\newcommand{\ga}{\gamma}

\def\bTheta{\mbox{\boldmath$\Theta$}}
\def\btheta{\mbox{\boldmath$\theta$}}

\def\bxi{\mbox{\boldmath$\xi$}}
\def\bX{\mbox{\boldmath$X$}}
\def\bx{\mbox{\boldmath$x$}}

\def\baselinestretch{1.7}
\usepackage{bm}
\usepackage{epsfig}

\begin{document}
\title{\vspace*{-1.5cm}
MultiLCIRT: An R package for multidimensional latent class item response models}
\author{Francesco Bartolucci\footnote{Department of Economics, Finance and Statistics,
University of Perugia, Via A. Pascoli, 20, 06123 Perugia.}
\footnote{{\em email}: bart@stat.unipg.it} ,
Silvia Bacci$^*$\footnote{{\em email}: silvia.bacci@stat.unipg.it} ,
Michela Gnaldi$^*$\footnote{{\em email}: gnaldi@stat.unipg.it}}  \maketitle
\def\baselinestretch{1.3}
\begin{abstract}\noindent
We illustrate a class of Item Response Theory (IRT) models for binary and ordinal 
polythomous items and we describe an \textsf{R} package for dealing with these
models, which is named \verb'MultiLCIRT'. The models at issue 
extend traditional IRT models allowing for ({\em i})
 multidimensionality and ({\em ii}) discreteness of latent traits. This class of models also allows for different parameterizations for the conditional distribution of the response variables given the latent traits, depending on both the type of link function and the constraints imposed on the discriminating and the difficulty item parameters. We illustrate how the proposed class of models may be estimated by the maximum likelihood approach via an Expectation-Maximization algorithm, which is implemented in the \verb'MultiLCIRT' package, 
and we discuss in detail issues related to model selection. In order to illustrate this package, we analyze two datasets: one concerning binary items and referred to the measurement of ability in mathematics and the other one coming from the administration of ordinal polythomous items for the assessment of anxiety and depression. In the first application, we  illustrate how aggregating items in homogeneous groups through a model-based hierarchical clustering procedure which is implemented in the proposed package. In the second application, we describe the steps to select a specific model having the best fit in
our class of IRT models.

\noindent \vskip5mm \noindent {\sc Keywords: EM algorithm, HADS data, model-based clustering, NAEP data.}
\end{abstract}
\newpage

\section{Introduction}\label{sec:introduction}

Item Response Theory (IRT) models are commonly used to measure latent traits, through the analysis of data
deriving from the administration of questionnaires made of items
with dichotomous or polythomous responses (also known, in the
educational setting, as dichotomously or polythomously-scored items). Usually, traditional IRT models are based  on the unidimensionality assumption, which means that all items contribute to measure the same latent trait. Moreover, 
in some cases, the normality assumption of this latent trait is explicitly introduced. Unfortunately, in several practical situations both of these assumptions are restrictive. 
Therefore, several extensions of traditional IRT models have been proposed in the
literature in order  to make the models more flexible and realistic. 

Firstly,
some authors dealt with multidimensional extensions of IRT models to take into account
that questionnaires are often designed to measure more than one latent trait. Among the
main  contributions, we remind
\cite{dun:ste:87}, \cite{agr:93}, and \cite{kel:rij:94}, who proposed a number of
examples of loglinear multidimensional IRT models, \cite{kel:96} for a multidimensional
version of the Partial Credit Model \citep{mast:82}, and \cite{ada:wil:wan:97} for a wide class of Rasch type
\citep{rasch:60, wri:mas:82} extended  models;
see \cite{rec:09} for a thorough overview of this topic.

Another advance in the IRT literature
concerns the assumption that the population
under study is composed by homogeneous classes of individuals who have very similar
latent characteristics  \citep{laza:henr:68,good:74}.
In some contexts, where the aim is clustering individuals, this is a convenient
assumption; in health care, for instance, by introducing this assumption
we single out
a certain number of clusters of patients receiving the same clinical treatment.
Secondly, this assumption allows us to estimate the
model in a semi-parametric way, namely
without formulating any assumption on the latent trait distribution. 
Moreover, it is possible to implement the maximum marginal likelihood method making use of the Expectation-Maximization (EM) algorithm \citep{demp:lair:rubi:77}, skipping in this way the problem of intractability of multidimensional integral which characterizes the marginal likelihood when a continuous distribution is assumed for the latent variable. 
At this regards, \cite{chr:02} outlined, through a simulation study, the computational problems encountered during the estimation process of a multidimensional model based on a multivariate normally distributed ability.   
See also \cite{mas:85}, \cite{lan:rost:88}, \cite{hei:96}, and 
\cite{for:07} for a comparison
between traditional IRT models
with those formulated by a latent class approach.
For some examples of discretized variants of IRT models we also
remind 
\cite{lind:91}, \cite{for:92}, \cite{hoi:mol:97}, \cite{ver:01}, and \cite{smit:03}. Another interesting
example of
combination between the IRT approach and 
latent class approach is represented by the mixed Rasch
model for binary and ordinal polythomous data 
 \citep{rost:90, rost:91, dav:rost:95},  builded as a mixture of latent classes with a separate Rasch model assumed to hold within each of these classes.

As concerns the combination of the two above mentioned extensions,
in the context of dichotomously-scored items \cite{bart:07}
proposed a class of
multidimensional latent class (LC) IRT models, where:
({\em i}) more latent traits are simultaneously considered 
and each item is associated with only one of them
(between-item multidimensionality; for details see \cite{ada:wil:wan:97} and 
\cite{zhang:04}) and ({\em ii}) these latent traits are represented by a
random vector with a discrete distribution common to all subjects
(each support point of such a distribution identifies a different
latent class of individuals). Moreover, in this class of models
either a Rasch \citep{rasch:60} or a
two-parameter logistic (2PL) parameterization \citep{birn:68} may be adopted for
the probability of a correct response to each item. 
Similarly to \cite{bart:07}, \cite{dav:08} proposed the diagnostic model, which, as main difference, assumes fixed rather than free abilities. An interesting comparison of multidimensional IRT models based on continuous and discrete latent traits was performed by \cite{hab:dav:lee} in terms of goodness of fit, similarity of parameter estimates and computational time required.  

The aim of the present paper is to illustrate a wide class of multidimensional LC IRT models that accommodates both binary and ordinal polythomous items in the same framework and
the \textsf{R}  package \verb'MultiLCIRT' to deal with these models \citep{multi:pack}.
The class of model at issue includes that proposed by \cite{bart:07} as a special case and is
formulated so that different parameterizations may be adopted for the conditional
distribution of the response variables, given the latent traits. We mainly refer to the
classification
criterion proposed by  \cite{mol:83}; see also \cite{agr:90} and \cite{ark:01}.
Relying on the
type of link function, it allows to discern among: ({\em i}) graded response
models, based on global (or cumulative) logits and ({\em ii}) partial credit models, which
make use of
local (or adjacent category) logits. For each of these link functions, 
we explicitly consider the possible
presence of constraints on the discriminating and the difficulty item parameters. As
concerns the first element, we take into account the possibility that all items have the
same discriminating power against the possibility that they discriminate differently.
Moreover, we discern the case in which each item differs from the others for
different distances  between the difficulties of consecutive response categories
and the special case in
which the distance between difficulty levels from category to category
is the same for
all items. On the basis of the choice of all the mentioned features
(i.e., type of link function, item discriminating parameters, and item difficulties),
different item
parameterizations are defined. We show how these parameterizations
result in an extension of traditional IRT models, by introducing assumptions of
multidimensionality and discreteness of latent traits.

In order to estimate each model in the proposed class,
we rely on an EM algorithm.
Moreover, special attention is given to the model selection procedure, that aims 
at choosing the optimal number of latent classes, the type of link function,  the
parameterization for the item discriminating and difficulty parameters, and the
number of latent dimensions and the
allocation of items within each dimension. 
Algorithms for the estimation of the proposed class of models have been implemented in the
package \verb'MultiLCIRT' \citep{multi:pack}. We describe the main functions of
this package, with special attention to input required and output supplied. 

Several other \textsf{R}  packages are suitable for estimating IRT models and their extensions, but to our knowledge neither of them treat  multidimensionality and discreteness of latent traits at the same time. Among others, we remind packages \verb'mirt' \citep{mirt:pack}, \verb'plink' \citep{plink:pack}, \verb'irtProb' \citep{irtProb:pack} (only for binary items), and \verb'plRasch' \citep{plRasch:pack} (only for Rasch-type models), for the estimation of multidimensional IRT models. Moreover, we cite package \verb'mixRasch' \citep{mixRasch:pack}  for the discreteness assumption of latent trait (only for Rasch-type models). 

In order to illustrate the proposed class of models and the use of \verb'MultiLCIRT' package, we describe two applications. The first one is about  a dataset extrapolated by a larger dataset collected in 1996 by the Educational Testing Service within the National Assessment of Educational Progress (NAEP) project.  Following \cite{bart:07}, we illustrate how performing a hierarchical model-based clustering of a set of binary items in homogeneous groups, each of them measuring different latent traits. The second application concerns
a dataset collected by a set of ordinal polythomous items on anxiety and depression of
oncological patients, and formulated following the ``Hospital
Anxiety and Depression Scale'' (HADS) developed by \cite{zig:sna:1983}. Through this
application, the steps of the model selection procedure are illustrated and the
characteristics of each latent class, in terms of estimated levels of the latent traits,
are described with reference to the selected model.

The reminder of this paper is organized as follows.   In Section
\ref{sec:model} we describe the proposed class of multidimensional LC IRT models and we focus on the different types of parameterizations that can be specified. Section \ref{sec:inference} is devoted to maximum likelihood estimation and to model selection.
Section \ref{sec:package} is devoted to illustrate functions implemented in \verb'MultiLCIRT' package.
In Section \ref{sec:application}, the proposed class of  models is illustrated through
the analysis of two real datasets with specific reference to \verb'MultiLCIRT' package,
whereas some final remarks are reported in Section \ref{sec:conclusion}.

\section{The class of multidimensional latent class IRT models}\label{sec:model}

In the following, we describe the proposed class of models by illustrating the different specifications due to the type of link function and the constraints on the item parameters. Then, in order to efficiently implement the parameter estimation, the formulation of the class of models is described in matrix notation.

\subsection{The model}

Let $X_j$ denote the response variable for the $j$-th item of the questionnaire, with $j =1, \ldots, r$. We assume that each item $j$ has $l_j$ categories, indexed by $x$ from 0 to $l_j-1$, where $l_j = 2$ corresponds to the case of a binary item ($x = 0, 1$).
Let $s$ be the number of different latent traits measured by the items,
let $\bTheta = (\Theta_1,\ldots, \Theta_{s})\tr$ be a vector of latent variables
corresponding to these latent traits, and let $\btheta=
(\theta_1,\ldots,\theta_s)\tr$ denote one of its possible realizations. The random vector $
\bTheta$ is assumed to have a discrete distribution with $k$ support points, denoted by
$\bxi_1,\ldots,\bxi_k$, and probabilities $\pi_1,\ldots,\pi_k$, with $\pi_c = p(\bTheta=
\bxi_c)$.
Moreover, let $\delta_{jd}$ be a dummy variable equal to $1$ if item $j$
measures latent trait of type $d$ and to 0 otherwise, with
$j=1,\ldots,r$ and $d=1, \ldots, s$. We also denote the
conditional response probability that a subject with latent traits (or abilities) levels given by $\b\th$ responds by category $x$ to item $j$ as follows
\[
\la_{jx}(\b\th)=p(X_j=x|\b\Th=\b\th),\quad x=0,\ldots,l_j-1,
\]
and we let $\b\la_j(\b\th)=(\la_{j0}(\b\th),\ldots,\la_{j,l_j-1}(\b\th))\tr$, the elements of which sum up to 1.\newpage

The IRT models that are here of interest may be expressed through the following general formulation
\begin{equation}\label{eq:poly_gen2}
g_x(\b\la_j(\b\th)) = \ga_j (\sum_{d=1}^{s} \delta_{jd} \theta_d - \beta_{jx}),
\quad j=1,\ldots,r,\:
x=1,\ldots,l_j-1,
\end{equation}
where $g_x(\cdot)$ is a link function specific of category $x$ and  $\ga_j$ and $\be_{jx}$ are item parameters, usually identified as  discriminating and difficulty indices and on which suitable constraints need to be assumed. 

The formulation of each model belonging to the proposed class
depends on the specification on the following three elements: 
\begin{enumerate}
\item {\em Type of link function}: we consider the link based on: ({\em i}) global (or
cumulative) logits and ({\em ii}) local (or adjacent categories) logits.
In the first case, the link function is defined as
\[
g_x[\b\la_j(\b\th)]
=\log\frac{\la_{x|\th}^{(j)}+\cdots+\la_{l_j-1|\th}^{(j)}}
{\la_{0|\th}^{(j)}+\cdots+\la_{x-1|\th}^{(j)}}
=\log\frac{p(X_{j}\geq x|\theta)}{p(X_{j} < x|\theta)},\quad x=1, \ldots, l_j-1,
\]
and compares the probability that item response is in category $x$ or higher with
the probability that it is in a lower category. Moreover, with local
logits we have that
\[
g_x[\b\la_j(\th)]
=\log\frac{\la_{x|\th}^{(j)}}{\la_{x-1|\th}^{(j)}}
=\log\frac{p(X_{j}=x |\theta)}{p(X_{j}=x-1|\theta)},
\quad x=1,\ldots,l_j-1,
\]
and then the probability of each category $x$ is compared with the probability of the
previous category; in the context of IRT models, see also \cite{forc:bart:2004}.
We note that, in case of binary items, the two types of logits coincide.  
%
%
%
Global logits are typically used when the trait of interest is assumed to be continuous
but latent, so that it can be observed only when each subject reaches a given threshold
on the latent continuum. On the contrary, local logits are used to identify one or more
intermediate levels of performance on an item and to award a partial credit for reaching
such intermediate levels. 
IRT models based on global logits are also known as graded response models, those based
on local logits are known as partial credit models. 

\item {\em Constraints on the discriminating parameters}: we consider: ({\em i})
a general
situation in which each item may discriminate differently from the others and 
({\em ii}) a special case in which all the items discriminate in the same way,
that is
\begin{equation}
\gamma_j=1,\quad j = 1, \ldots, r.\label{eq:same_discrimination}
\end{equation}
Note that, in both cases, we assume that,
within each item, all $l_j > 2$ response categories share the same $\gamma_j$, in order to keep
the conditional probabilities away
from crossing and so avoiding degenerate
conditional response probabilities. We also observe that, with binary items, case ({\em i}) corresponds to a two-parameters logistic (2PL) formulation \citep{birn:68}, whereas case ({\em ii}) is referred to a Rasch parameterization \citep{rasch:60}.  
\item {\em Formulation of item difficulty parameters}:
limited to the case of polythomously-scored items (i.e., $l_j > 2$), we consider:
({\em i}) a general situation in which the parameters $\be_{jx}$ are unconstrained
and ({\em ii}) a special case in which these parameters are constrained so that
the distance between difficulty levels from category to category is the same for
each item (rating scale parameterization). Obviously, the second case
makes sense when all items have the same number of response categories,
that is $l_j=l$, $j=1,\ldots,r$. This constraint may be expressed as
\begin{equation}
\beta_{jx} = \beta_j + \tau_x,\quad j=1,\ldots,r,\: x = 0,\ldots,l-1,
\label{eq:rating_scale}
\end{equation}
where $\beta_j$ indicates the difficulty of item $j$ and $\tau_x$ is the difficulty of
response category $x$ for all $j$. Note that, with binary items,
we have $\beta_{jx} = \beta_{j}$. 
\end{enumerate}

By combining the above constraints, we obtain four different
specifications of the item parametrization (second member of eqn. \ref{eq:poly_gen2}), based on free or constrained discriminating
parameters and on a rating scale or a free parameterization for difficulties (Table \ref{tab:multi_h}).

\begin{table}[!ht]\centering\vspace*{0.5cm}
{\small
\begin{tabular}{ccc}
\hline \hline
discriminating    &  difficulty & item \\
indices & levels  &             parameterization           \\
\hline
    free     &        free       &    $\gamma_j (\sum_d \delta_{jd} \theta_d - \beta_{jx})$\\
    free     &   constrained &     $\gamma_j [\sum_d \delta_{jd} \theta_d - (\beta_{j} + \tau_x)]$\\
    constrained   &   free    &      $ \sum_d \delta_{jd} \theta_d - \beta_{jx}$  \\
    constrained   &   constrained &      $\sum_d \delta_{jd} \theta_d - (\beta_{j} + \tau_x)$  \\
    \hline
\end{tabular}}
\caption{\em Specifications of the item parametrization}  \label{tab:multi_h}
\vspace*{0.5cm}\end{table}
\newpage
According to the above illustrated item parameterization and also according to the type of link function, several different types of
multidimensional LC IRT models derive which extend to multidimensionality and discreteness of latent traits some well-known traditional unidimensional IRT models. For instance, we may define the multidimensional LC Graded Response Model (GRM), which is an extension  of the GRM by \cite{sam:69},  as
\begin{equation}\label{eq:multi_LC_GRM}
\log \frac{p(X_{j}\geq x|\bTheta=\btheta)}{p(X_{j} < x|\bTheta = \btheta)}=
\gamma_j (\sum_{d=1}^{s} \delta_{jd} \theta_d - \beta_{jx}),\quad x=1, \ldots, l_j-1,
\end{equation}
and  the multidimensional LC Rating Scale Model (RSM), which is an extension  of the RSM by \cite{andr:78},  as 
\begin{equation}\label{eq:multi_LC_RSM}
\log \frac{p(X_{j}=x |   \bTheta = \btheta)}{p(X_{j}=x-1|   \bTheta = \btheta)}  =
\sum_{d=1}^{s} \delta_{jd} \theta_d - (\beta_{j} + \tau_{x}), \quad x=1,
\ldots, l-1.
\end{equation}
Note that when $l_j=2$, $j=1,\ldots,r$, so that item responses
are binary, equations (\ref{eq:multi_LC_GRM}) and (\ref{eq:multi_LC_RSM}) specialize,
respectively, in the multidimensional LC 2PL model and in the multidimensional LC Rasch
model, both of them described by \cite{bart:07}.

In all cases, the discreteness of the distribution of the random vector $\bTheta$
implies that the manifest distribution of $\b X=(X_1,\ldots,X_r)\tr$
for all subjects in the $c$-th latent class is equal to
\begin{equation}
p(\b x)=p(\bX = \bx) = \sum_{c=1}^{k} p(\bX = \bx|\bTheta=\bxi_c) \pi_c,
\label{eq:prob_mani}
\end{equation}
where, due to the 
classical assumption of {\em local independence}, we have
\begin{eqnarray}
p(\b x|c)=
p(\bX = \bx|\bTheta=\bxi_c) & = & \prod_{j=1}^{r} p(X_{j} = x_j|\bTheta=\bxi_c) =
\nonumber\\
                            & = & \prod_{d=1}^{s} \: \prod_{j\in\cgl J_d}
                            p(X_j = x_j|\Theta_d=\xi_{cd}), 
                            \label{eq:prob_cond}
\end{eqnarray}
where $\cg J_d$ denotes the subset of $\cg J=\{1,\ldots,r\}$ containing the indices
of the items measuring the $d$-th latent trait, with $d=1,\ldots,s$
and $\xi_{cd}$ denoting the $d$-th elements of $\b\xi_c$.

In order to ensure the identifiability of the proposed models, 
suitable constraints on the 
parameters are required. With reference to the general equation (\ref{eq:poly_gen2}),
we  require that, for each latent trait, one discriminating index  is equal to 1 and one
difficulty  parameter  is equal to 0.
More precisely, let $j_d$ be a specific element of $\cg J_d$, say the first.
Then, when the discriminating indices are not constrained to be constant
as in (\ref{eq:same_discrimination}), we assume that
\[
\ga_{j_d} = 1,\quad d=1,\ldots,s.
\]
Moreover, with free item difficulties we assume that
\begin{equation}
\be_{j_d1} = 0,\quad d=1,\ldots,s, 
\label{eq:constraing_beta1}
\end{equation}
whereas with a rating scale parameterization based on (\ref{eq:rating_scale}),
we assume
\begin{equation}
\be_{j_d} = 0,\quad d=1,\ldots,s,\quad\mbox{and}\quad\tau_1=0. 
\label{eq:constraing_beta2}
\end{equation}

Coherently with the mentioned identifiability constraints, the number of free parameters
of a multidimensional LC IRT model is obtained by summing the
number of free probabilities $\pi_c$, the number of ability
parameters $\xi_{cd}$, the number of free item difficulty parameters $\beta_{jx}$, and
that of free item discriminating parameters $\ga_j$.
We note that the number of free parameters does not depend on the
type of logit, but only on the type of parametrization assumed 
on item discriminating and
difficulty parameters, as shown in Table \ref{tab:num_par}. In any case, the number of
free class probabilities is equal to $k-1$ and the number of ability
parameters is equal to $sk$. However, the number of free item difficulty parameters
is given by $[\sum_{j=1}^r(l_j-1)-s]$ under an unconstrained difficulties
parameterization, by $[(r-s)+(l-2)]$ under a rating scale
parameterization, and by $r-s$ in case of binary items ($l_j = 2$, $j = 1, \ldots, r$).
Finally, the number of free item discriminating parameters is equal
to $(r-s)$ under an unconstrained discriminating parameterization, being 0 otherwise.\\ 

\begin{table}[!ht]\centering\vspace*{0.5cm}
{\small
\begin{tabular}{cc|l}
\hline \hline
discriminating    &  difficulty & \multicolumn1c{Number of free parameters} \\
indices & levels  &          \multicolumn1c{$(\#{\rm par})$}              \\\hline
    free     &        free       &  $(k-1) + sk + \big[\sum_{j=1}^r(l_j-1)-s\big] + (r-s)$                          \\
    free     &   constrained & $(k-1) + sk + [(r-s) + (l-2)] + (r-s)$         \\
        constrained   &   free     &  $(k-1) + sk + \big[\sum_{j=1}^r(l_j-1)-s\big]$                                             \\
    constrained   &   constrained  & $(k-1) + sk + [(r-s) + (l-2)]$                                 \\
\hline
\end{tabular}}
\caption{\em Number of free parameters for different constraints on item discriminating and difficulty parameters.}  \label{tab:num_par}\vspace*{0.5cm}
\end{table}

\subsection{Formulation in matrix notation}
In order to efficiently implement parameter estimation,
in this section we express the above described class of
models by using the matrix notation. For simplicity,
we consider the case in which every item has the same number of response
categories, that is $l_j=l$, $j=1,\ldots,r$;
the extension to the general case
in which items may also have a different number of response categories is
straightforward. In the following, by $\b 0_a$ we denote
a column vector of $a$ zeros, by $\b O_{ab}$ an $a\times b$ matrix of zeros, by
$\b I_a$ an identity matrix of size $a$, by $\b 1_a$ a column vector of $a$ ones.
Moreover, we use the symbol $\b u_{ab}$ to denote a column vector of $a$ zeros with
the $b$-th element equal to one and
$\b T_a$ to denote an $a\times a$ lower triangular matrix of ones.
Finally, by $\ot$ we indicate the Kronecker product.

As concerns the link function in (\ref{eq:poly_gen2}), it may be expressed in a
general way to include different types of parameterizations
\citep{glon:mccu:95, colo:forc:01} as follows:
\begin{equation}\label{eq:logits}
\b g[\b\la_j(\b\th)] = \b C\log[\b M\b\la_j(\b\th)],
\end{equation}
where the vector $\b g[\b\la_j(\b\th)]$ has elements
$g_x[\b\la_j(\b\th)]$ for $x=1,\ldots,l-1$. Moreover, $\b C$ is a matrix
of constraints of the type
\[
\b C = (\begin{matrix}-\b I_{l-1}
& \b I_{l-1}\end{matrix}),
\]
whereas, for the global logit link, matrix $\b M$ is equal to
\[
\b M = \begin{array}{ll}
\left(\begin{matrix}
\b T_{l-1} & \b 0_{l-1} \\
\b 0_{l-1} & \b T_{l-1}\tr
\end{matrix}\right),
\end{array}
\]
and for the local logit link it is equal to
\[
\b M = \begin{array}{ll}
\left(\begin{matrix}
\b I_{l-1} & \b 0_{l-1} \\
\b 0_{l-1} & \b I_{l-1}
\end{matrix}\right).
\end{array}
\]
How to obtain the probability vector $\b\la_j(\b\th)$ on the basis of a vector
of logits defined as in (\ref{eq:logits}) is described in \cite{colo:forc:01},
where a method to compute the derivative of a suitable vector of canonical
parameters for $\b\la_j(\b\th)$ with respect to these logits may be found.

Once the ability and difficulty parameters are included in the single vector
$\b\phi$ and taking into account that the distribution of $\b\Th$ has $k$ support
points, assumption (\ref{eq:poly_gen2}) may be expressed through the general formula
$$
\b g[\b\la_j(\b\xi_c)] =
\ga_j\b Z_{cj}\b\phi,\quad c=1,\ldots,k,\:j=1,\ldots,r,
$$
where $\b Z_{cj}$ is a suitable design matrix. The structure of the
parameter vector $\b\psi$ and of these design matrices depend on the type of
constraint assumed on the difficulty parameters, as we explain below.

When the difficulty parameters are unconstrained, $\b\phi$ is a
column vector of size $sk+r(l-1)-s$, which is obtained from
$$
(\xi_{11},\ldots,\xi_{1s},\ldots,\xi_{ks},\beta_{11},\ldots,
\beta_{1,l_1-1}, \ldots, \beta_{r,l_r-1})\tr
$$
by removing the parameters constrained to be 0; see constraint (\ref{eq:constraing_beta1}).
Accordingly, for $c=1,\ldots,k$ and $j=1,\ldots,r$, the design matrix
$\b Z_{cj}$ is obtained by removing suitable columns from the matrix
\[
\begin{pmatrix}
\b 1_{l-1}(\b u_{kc}\ot\b u_{sd})\tr & \b u_{rj}\tr\ot\b I_{l-1}
\end{pmatrix},
\]
where $d$ is the dimension measured by item $j$.
On the other hand, under a rating scale parameterization, $\b\phi$ is a vector of
size $sk+(r-s)+(l-2)$ which is obtained from
$$
(\xi_{11},\ldots,\xi_{1s},\ldots,\xi_{ks},\beta_1,
\ldots,\beta_r,\tau_1,\ldots,\tau_{l-1})\tr
$$
by removing the parameters constrained to be 0 in (\ref{eq:constraing_beta2}).
Accordingly, the design matrix $\b Z_{cj}$ is obtained by removing
specific columns from
\[
\begin{pmatrix}
\b 1_{l-1}(\b u_{kc}\ot\b u_{sd})\tr & \b 1_{l-1}\b u_{rj}\tr & \b I_{l-1}
\end{pmatrix},
\]
where, again, $d$ is the dimension measured by item $j$.

\section{Likelihood inference}\label{sec:inference}
In this section, we deal with  likelihood inference for the models proposed in the previous section. In particular, we first show how to compute the model log-likelihood and how to maximize it by an EM algorithm. Finally, we deal with some relevant aspects of model selection, mainly concerning the assessment of the latent dimensions.
\subsection{Maximum likelihood estimation}
On the basis of an observed sample of dimension $n$,
the log-likelihood of a model formulated as proposed in Section
\ref{sec:model} may be expressed as
\[
\ell(\b\eta)=\sum_{\bl x} n_{\bl x}\log[p(\b x)],
\]
where $\b\eta$ is the vector containing all the free model parameters,
$n_{\bl x}$ is the frequency of the response configuration $\b x$,
$p(\b x)$ is computed according to (\ref{eq:prob_mani}) and (\ref{eq:prob_cond})
as a function of $\b\eta$, and by $\sum_{\bl x}$ we mean
the sum extended to all the possible response configurations $\b x$.

In oder to maximize $\ell(\b\eta)$ with respect to $\b\eta$ we
use an EM algorithm \citep{demp:lair:rubi:77} that is implemented
in a similar way as described in \cite{bart:07}, to which we refer
for some details. First of all, denoting by $m_{c,\bl x}$ the (unobserved)
frequency of the response configuration $\b x$ and the latent
class $c$, the {\em complete} log-likelihood is equal to
\begin{equation}
\ell^*(\b\eta)=\sum_c\sum_{\bl x}m_{c,\bl x}\log[p(\b x|c)\pi_c].
\label{eq:comp_lk}
\end{equation}
Now we denote by $\b\eta_1$ the subvector of $\b\eta$
which contains the free latent class probabilities
and by $\b\eta_2$ the subvector containing the remaining free parameters.
More precisely, we let $\b\eta_1=\b\pi$, with $\b\pi=(\pi_2,\ldots,\pi_k)\tr$,
and $\b\eta_2=(\b\ga\tr,\b\phi\tr)\tr$, where $\b\ga$ is obtained by removing
from $(\ga_1,\ldots,\ga_r)\tr$ the parameters which are constrained to be
equal to 1 to ensure identifiability.
Obviously, $\b\ga$ is not present when constraint
(\ref{eq:same_discrimination}) is adopted. Then, we can decompose the complete log-likelihood as
\[
\ell^*(\b\eta)=\ell_1^*(\b\eta_1)+\ell_2^*(\b\eta_2),
\]
with
\begin{eqnarray}
\ell_1^*(\b\eta_1)&=&\sum_c m_c\log\pi_c,\label{eq:comp_lk1}\\
\ell_2^*(\b\eta_2)&=&\sum_c\sum_j\b m_{cj}\tr\log\b\la_{cj},
\label{eq:comp_lk2}
\end{eqnarray}
where $m_c=\sum_{\bl x}m_{c,\bl x}$ is the number of subjects in latent class $c$
and $\b m_{cj}$ is the column vector with elements $\sum_{\bl x}I(x_j=x)m_{c,\bl x}$,
$x=1,\ldots,l_j-1$, with $I(\cdot)$ denoting the indicator function.

The EM algorithm alternates the following
two steps until convergence:
\begin{description}
\item[E-step:] compute the conditional expected value of $\ell^*(\b\eta)$
given the observed data and the current value of the parameters;
\item[M-step:] maximize the above expected value with respect to $\b\eta$,
so that this parameter vector results updated.

\end{description}

The E-step consists of computing, for every $c$ and $\b x$, the expected value
of $m_{c,\bl x}$ given $n_{\bl x}$ as follows
\[
\hat{m}_{c,\bl x} = n_{\bl x} \frac{p(\b x|c)\pi_c}{\sum_h p(\b
x|h)\pi_h}
\]
and then substituting these expected frequencies in (\ref{eq:comp_lk}). On the
basis of $\hat{m}_{c,\bl x}$ we can obtain the expected frequencies $\hat{m}_c$
and $\hat{\b m}_{cj}$ which, once substituted in (\ref{eq:comp_lk1}) and
(\ref{eq:comp_lk2}), allow us to obtain the expected values of $\ell_1^*(\b\eta_1)$
and $\ell_2^*(\b\eta_2)$, denoted by $\hat{\ell}_1^*(\b\eta_1)$ and
$\hat{\ell}_2^*(\b\eta_2)$, respectively.

At the M-step, the function obtained as described above is maximized with respect
to $\b\eta$ as follows. First of all, regarding the parameters in
$\b\eta_1$ we have
an explicit solution given by
\[
\pi_c= \frac{\hat{m}_c}{n},\quad c=2,\ldots,k, 
\]
which corresponds to the maximum of $\hat{\ell}^*_1(\b\eta_1)$.
To update the other parameters, we maximize $\hat{\ell}^*_2(\b\eta_2)$
by a Fisher-scoring algorithm that we illustrate in the following.

The Fisher-scoring algorithm alternates a step in which the parameter vector $\b\ga$
is updated with a step in which the parameter vector $\b\phi$ is updated.
The first step consists of adding to the current value of each
free $\ga_j$ the ratio $s_{2j}^*/f_{2j}^*$,
where $s_{2j}^*$ denotes the score for $\hat{\ell}^*_2(\b\eta_2)$
with respect to $\ga_j$ and $f_{2j}^*$ denotes the corresponding
information computed at the current value of the parameters. These have
the following expressions:
\begin{eqnarray*}
s_{2j}^* &=& \sum_c\sum_j(\b Z_{cj}\b\phi)\tr
\b R_{cj}\tr(\hat{\b m}_{cj}-\hat{m}_c\b\la_{cj}),\\
f_{2j}^* &=& \sum_c\hat{m}_c\sum_j(\b Z_{cj}\b\phi)\tr
\b R_{cj}\tr[\diag(\b\la_{cj})-\b\la_{cj}\b\la_{cj}\tr]\b R_{cj}(\b Z_{cj}\b\phi),
\end{eqnarray*}
where $\b R_{cj}$ is the derivative matrix of the canonical parameter vector
for $\b\la_{cj}$ with respect to the vector of logits in (\ref{eq:logits});
see \cite{colo:forc:01}.
Then, the parameter vector $\b\phi$ is updated by adding the quantity
$(\b F_2^*)^{-1}\b s_2^*$, where $\b s_2^*$ is the score vector for
$\hat{\ell}^*_2(\b\eta_2)$ with respect to $\b\phi$ and $\b F_2^*$ denotes the
corresponding information computed at the current parameter value, which
have the following expressions:
\begin{eqnarray*}
\b s_2^* &=& \sum_c\sum_j\ga_j\b Z_{cj}\tr
\b R_{cj}\tr(\hat{\b m}_{cj}-\hat{m}_c\b\la_{cj}),\\
\b F_2^* &=& \sum_c\hat{m}_c\sum_j\ga_j^2\b Z_{cj}\tr
\b R_{cj}\tr[\diag(\b\la_{cj})-\b\la_{cj}\b\la_{cj}\tr]\b R_{cj}\b Z_{cj}.
\end{eqnarray*}

As usual, we suggest to initialize the EM algorithm by a deterministic
rule and by a multi-start strategy based on random starting values which
are suitable generated. In this way we can deal with the multimodality of
the model likelihood.

\subsection{Model selection}\label{sec:modsel}
The formulation of a specific model in the class of multidimensional LC IRT models univocally depends on: ({\em i}) the number of latent classes ($k$); 
({\em ii}) the adopted parameterization in terms of link function $g_x(\cdot)$,  ({\em iii}) the
constraints on the item parameters
$\ga_j$ and $\beta_{jx}$, and ({\em iv}) the number ($s$) of latent dimensions and the
corresponding allocation of items within each dimension ($\de_{jd}$, $j=1,\ldots,r$,
$d=1,\ldots,s$).

Thus, the model selection implies the adoption of a number of choices, for each of the 
mentioned aspects, by using suitable criteria. We suggest to mostly rely 
on the likelihood ratio (LR) test to compare nested models and on the Bayesian information criterion  \cite[BIC;][]{sch:78} to obtain a relative measure
of lost information when a given model is used to describe observed data and, therefore, to compare non-nested models. More precisely, the LR test is here preferred to the Wald test, because it does not require to compute the information matrix of the model. Moreover, the BIC index has to be preferred to other information criteria, because it satisfies some nice properties. Mainly, under certain regularity
conditions it is asymptotically consistent \citep{keribin:2000}. Moreover,
since it applies a larger penalty for additional parameters (for reasonable
sample sizes) in comparison with other criteria, BIC tends to select a more parsimonious
model.

In summary, we suggest to base the selection of the number of latent classes and the type of logit link function on the BIC index, whereas the selection of the item discriminating and difficulty parameterization may be performed on the basis of the LR test.  

The LR test
may also be used to test the null hypothesis that  items in $\cg I_{d_1}$ and $\cg I_{d_2}$ measure the same latent trait.
Therefore,  being equal all the other elements of the model (number of latent classes, constraints on item parameters, type of logit), the general model with $s$ dimensions is compared with a 
restricted version with $s-1$ dimensions, where items in $\cg I_{d_1}$ and $\cg I_{d_2}$ are collapsed in the same group. The LR test statistic is given by $-2\cdot (\hat{\ell}_0 - \hat{\ell}_1)$, where $\hat{\ell}_0$ and $\hat{\ell}_1$ denote the maximum of log-likelihood of the restricted model and of the general model, respectively. Under H$_0$, the LR statistics is asymptotically distributed as a $\chi_q^2$, where $q$ is given by the difference in the number of parameters (see Table \ref{tab:num_par}) between the two nested models being compared.  

By repeating this LR test for dimensionality in a
suitable way, we can cluster items so that items in the same group
measure the same ability. On the basis of this principle,
\cite{bart:07} proposed a hierarchical clustering algorithm based on Wald test and not yet implemented in \textsf{R}: its implementation for LR test is supplied in the \verb'MultiLCIRT' package. This algorithm builds a sequence of nested models: the most
general one is that with a separate dimension for each item
(corresponding to the classic LC model) and the
most restrictive model is that with only one dimension common to all
items (uni-dimensional model). The clustering procedure performs
$r-1$ steps. At each step, the LR test statistic is computed for every pair of possible
aggregations of items (or groups of items). The aggregation with the
minimum value of the statistic (or equivalently the highest
$p$-value) is then adopted and the corresponding model fitted before
moving to the next step. 

The output of the above clustering algorithm may be displayed
through a dendrogram that shows the deviance between the initial
($k$-dimensional) LC model and the model selected at each step of
the clustering procedure. Obviously, the results of a cluster
analysis based on a hierarchical procedure depend on the adopted
rule to cut the dendrogram, which may be chosen according to several
criteria. We suggest to choose $\hat{s}= r-h+1$, where $h$ is the first step for that the $p$-value corresponding to the LR statistic  is smaller than a suitable threshold, such as 0.05.

\section{The MultiLCIRT package}\label{sec:package}
The class of multidimensional LC IRT models described in the previous sections may be estimated through a set of functions implemented in the \verb'MultiLCIRT' package of \textsf{R} software. In the following we illustrate the use of main functions; for related
details see the official documentation \citep{multi:pack}.
\subsection{Aggregating data}
To speed up  the estimation of the proposed models and the clustering of items we suggest to perform the analyses by aggregating original records having the same response pattern so as to obtain a matrix with a record for each distinct response configuration (rather than for each statistical unit). For this aim 
we may use function   \verb'aggr_data', which requires as input the data matrix of unit-by-unit response configurations. The following output is obtained: 
\begin{itemize}
\item \verb'data_dis' is the matrix of distinct configurations;
\item \verb'freq' is the vector of corresponding frequencies;
\item \verb'label' is the index of each provided response configuration among the distinct ones.
\end{itemize}
\subsection{Estimating a Multidimensional LC IRT model}
Parameters estimation  for multidimensional IRT models based on discreteness of latent traits is performed through function \verb'est_multi_poly', which requires the following main input:%
\begin{itemize}
\item \verb'S': matrix of all response sequences observed at least once in the sample and listed row-by-row. Usually, \verb'S' is matrix \verb'data_dis' obtained by applying function \verb'aggr_data' to original data.  Missing responses are allowed and they are coded as 999;
\item \verb'yv': vector of the frequencies of every response configuration in \verb'S', corresponding to output \verb'freq' by function \verb'aggr_data';
\item \verb'k': number of latent classes;
\item \verb'start': method of initialization of the algorithm:  0 for deterministic starting values,  1 for random starting values, and  2 for
arguments given by the user. In case of \verb'start = 2', we also need to specify as input the initial values of weights, support points, discriminating and difficulty item parameters;
\item \verb'link': type of link function: 0 for the standard LC model (i.e., no link function is specified), 1 for global logits, and 2 for local logits. In case of binary items, it is the same to specify \verb'link=1' or \verb'link=2';
\item \verb'disc': indicator of constraints on the discriminating item parameters:  0 if $\gamma_j = 1, \; j= 1,\ldots,r$ and 1 if $\gamma_j \neq 1$ for at least one $j= 1,\ldots,r$;
\item \verb'difl': indicator of constraints on the difficulty item parameters:  0 if difficulties are free and 1 if $\beta_{jx} = \beta_j + \tau_x$;
\item \verb'multi': matrix with a number of rows equal to the number of dimensions and elements in each row equal to the indices of the items measuring the dimension corresponding
to that row.
\end{itemize}
  
Function \verb'est_multi_poly' supplies the following output:
\begin{itemize}
\item \verb'piv': estimated vector of weights of the latent classes;
\item \verb'Th': estimated matrix of ability levels (support points) for each dimension and latent class;
\item \verb'Bec': estimated vector of difficulty item parameters (split in two vectors if \verb'difl=1');
\item \verb'gac': estimated vector of discriminating item parameters;
\item \verb'fv': vector indicating the reference item chosen for each latent dimension;
\item \verb'Phi': array of the conditional response probabilities for every item and latent class;
\item \verb'Pp': matrix of the posterior probabilities for each response configuration and latent class;
\item \verb'lk': log-likelhood at convergence of the EM algorithm;
\item \verb'np': number of free parameters;
\item \verb'aic': Akaike Information Criterion index \citep{aka:73};
\item \verb'bic': Bayesian Information Criterion index \citep{sch:78}.
\end{itemize}

\subsection{Analysis of dimensionality}
As outlined in Section \ref{sec:modsel}, the analysis of dimensionality is based on two main procedures: ({\em i}) using a LR test to compare pairs of nested models that differ only by the way items are grouped and ({\em ii}) performing a model-based hierarchical clustering, when the grouping structure of items is far from clear.  Procedure ({\em i}) may be carried on by function \verb'test_dim' and procedure ({\em ii}) may be carried on by function \verb'class_item', being both of them based on the above described \verb'est_multi_poly' function. Both of these functions rely on the same input than  \verb'est_multi_poly', with just some differences. 

Function \verb'class_item' does not require any specification of multidimensionality structure, being it the object of analysis, whereas  function \verb'test_dim' requires to specify the multidimensional structure of both nested models, through the following arguments: 
\begin{itemize}
\item \verb'multi0': 	 matrix specifying the multidimensional structure of the restricted model, being the unidimensionality the default;
\item \verb'multi1': matrix specifying the multidimensional structure of the larger model.
\end{itemize}

Output of \verb'test_dim' consists in the following elements:
\begin{itemize}
\item \verb'out0': 	output for the restricted model obtained from \verb'est_multi_poly';
\item \verb'out1': 	output for the larger model obtained from \verb'est_multi_poly';
\item \verb'dev': 	LR test statistic;
\item \verb'df': number of degrees of freedom for the LR test;
\item \verb'pv': $p$-value for the LR test.
\end{itemize}

Finally, output of \verb'class_item' is given by:
\begin{itemize}
\item \verb'merge':	list of items and/or groups of items collapsed at each step of the clustering procedure;
\item \verb'order':	list of items sorted in a suitable way to represent the hierarchy of clusters through a dendrogram;
\item \verb'height':	values of LR statistic obtained at each step of the clustering procedure;
\item \verb'lk	': maximum log-likelihood of the model resulting from each aggregation;
\item \verb'np':	 number of free parameters of the model resulting from each aggregation;
\item \verb'groups':	 list of groups resulting (step-by-step) from the hierarchical clustering.
\end{itemize}
Elements \verb'merge', \verb'order', and \verb'height' can be used as input to plot a dendrogram.

\section{Examples}\label{sec:application}
In the following, we describe the use of \verb'MultiLCIRT' package through two applications on different datasets. Firstly, we illustrate the hierarchical clustering procedure of a set of binary items. Then, the model selection problem is taken into account by considering a set of ordinal polythomous items and we illustrate how choosing the number of latent classes, the type of logit, and the constraints on item parameters. We also deal with 
dimensionality testing.


\subsection{Hierarchical clustering for NAEP data items}
We illustrate the hierarchical clustering procedure by using a dataset containing the responses of a sample of 1510 examinees to 12 binary items on mathematics. It has been extrapolated \citep{bart:for:05} from a larger dataset collected in 1996 by the Educational Testing Service within the National Assessment of Educational Progress (NAEP) project. Here we replicate the example of \cite{bart:07} performed on the same data, but based on the Wald test rather than the LR test and implemented in Matlab.

We start loading the data:
\begin{output}
> data(naep)
> X = as.matrix(naep)
\end{output}
Then, we aggregate our data through function \verb'aggr_data', so reducing the number of records from 1510 to 740: 
\begin{output}
> out1 = aggr_data(X) 
> S = out1$data_dis
> yv = out1$freq
> nrow(X)
1510
> nrow(S)
704
\end{output}

\begin{figure}[ht!]
   \centering
   \includegraphics[width=10cm]{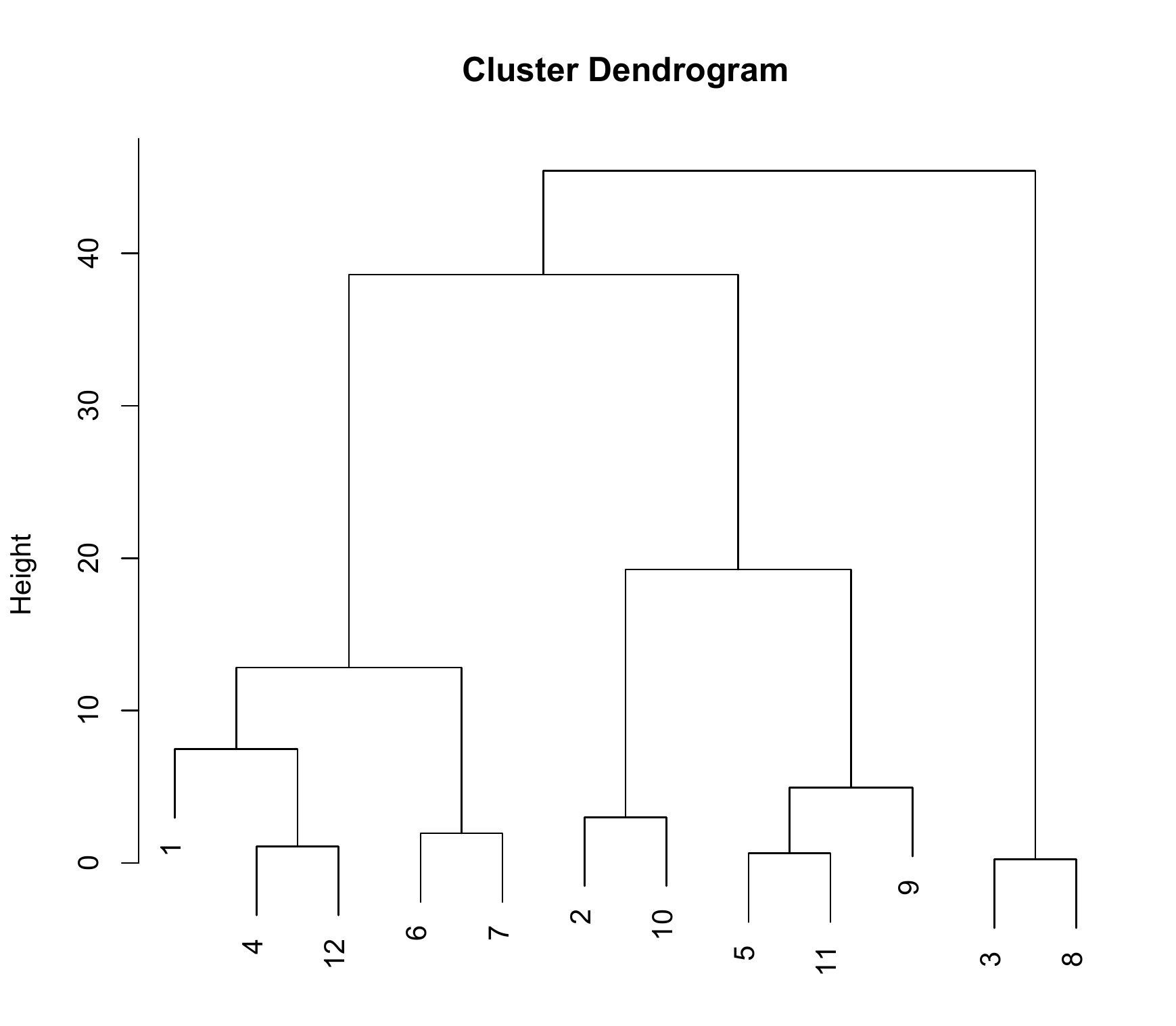}
   \caption{\emph{Dendrogram for the NAEP data.}}
   \label{fig:dendr}
\end{figure}

The hierarchical clustering of 12 items is performed through function \verb'class_item' and, according to \cite{bart:07}, it is based on the 2PL model with $k=4$ latent classes: 
\begin{output}
> out2 = class_item(S, yv, k=4, link=1, disc=1)  # it is the same with link = 2
> class(out2) = "hclust"
\end{output}
Results of clustering are displayed in the dendrogram obtained by command
\begin{output}
> plot(out2)
\end{output}
and shown in Figure \ref{fig:dendr}.

Detailed results are obtained by calling outputs in object \verb'out2' as
\begin{output}
> cbind(out2$merge, out2$height)

    [,1] [,2]       [,3]
 [1,]   -3   -8  0.2433178
 [2,]   -5  -11  0.6306477
 [3,]   -4  -12  1.0835975
 [4,]   -6   -7  1.9434207
 [5,]   -2  -10  3.0027518
 [6,]   -9    2  4.9478475
 [7,]   -1    3  7.4714241
 [8,]    4    7 12.8246846
 [9,]    5    6 19.2495712
[10,]    8    9 38.6127909
[11,]    1   10 45.4079171
\end{output} 

In the above output, the first two columns detect items (negative values) or groups (positive values)  collapsed at the corresponding step of the clustering procedure, whereas the third column identifies the corresponding LR statistic. More precisely, each positive value $h>0$ in column 1 and 2 denotes a specific group created at the $h$-th step of the procedure. For instance, the first aggregation is about items 3 and 8. These two items define group 1, as it will be called in the following. The second aggregation concerns items 5 and 11 defining group 2, and so on. Then, at sixth step item 9 is aggregated with items in group 2 (i.e, items 5 and 11); at seventh step item 1 is aggregated with items in group 3 (i.e., items 4 and 12), and so on. 
To visualize all groups we digit

  \begin{output}
> out2$group
\end{output}

To simplify the interpretation we may rearrange the results from the above call in Table \ref{tab:out2}, where third column identifies the clusters and fourth column the corresponding LR statistic;  for a sake of completeness we also add the $p$-values (last column), calculated on the basis of a $\chi_2^2$ distribution.  

\begin{table}[!ht]\centering\vspace*{0.5cm}
{\small
\begin{tabular}{cclccc}
\hline \hline
 Step $h$  &    $s$ & Clusters & LR statistics  & $p$-value   \\
\hline
1  & 11  & \{1\}, \{2\}, \{4\}, \{5\}, \{6\}, \{7\}, \{9\}, \{10\}, \{11\}, \{12\}, \{3, 8\} &  0.243   &   0.885  \\
2  & 10   &  \{1\}, \{2\}, \{4\}, \{6\}, \{7\}, \{9\}, \{10\}, \{12\}, \{5, 11\}, \{3, 8\} &  0.631   &     0.960 \\
3  & 9   &   \{1\}, \{2\}, \{6\}, \{7\}, \{9\}, \{10\}, \{4, 12\}, \{5, 11\}, \{3, 8\} &  1.084   &     0.982  \\
4  & 8  &  \{1\}, \{2\}, \{9\}, \{10\}, \{6, 7\}, \{4, 12\}, \{5, 11\}, \{3, 8\} &  1.943   &     0.983   \\
5  & 7  & \{1\}, \{9\}, \{2, 10\}, \{6, 7\}, \{4, 12\}, \{5, 11\}, \{3, 8\} &  3.003   &     0.981  \\
6  & 6  & \{1\}, \{2, 10\}, \{6, 7\}, \{4, 12\}, \{5, 9, 11\}, \{3, 8\} &  4.948   &     0.960  \\
7  & 5  &   \{2, 10\}, \{6, 7\}, \{1, 4, 12\}, \{5, 9, 11\}, \{3, 8\} &  7.471   &     0.915 \\
8  & 4  & \{2, 10\}, \{1, 4, 6, 7, 12\}, \{5, 9, 11\}, \{3, 8\} &  12.825   &     0.686  \\
9  & 3  &  \{2, 5, 9, 10, 11\}, \{1, 4, 6, 7, 12\}, \{3, 8\} &  19.250  &     0.377  \\
10   & 2  &  \{1, 2, 4, 5, 6, 7, 9, 10, 11, 12\}, \{3, 8\} &  38.613  &     0.007  \\
11  & 1   & \{1, 2, 3, 4, 5, 6, 7, 8, 9, 10, 11, 12\} &  45.408  &      0.002  \\
\hline
\end{tabular}}
\caption{\em Output of the hierarchical clustering algorithm for the NAEP data.}  \label{tab:out2}
\end{table}

We note that intermediate and final results of the clustering procedure, shown in  Table \ref{tab:out2}, are similar to those illustrated in \cite{bart:07}. More precisely, we can detect $s = 3$ groups of items ($\{2, 5, 9, 10, 11\}, \{1, 4, 6, 7, 12\}, \{3, 8\} $), corresponding to different latent traits.

\subsection{Model selection for HADS data}
We describe the model selection procedure and the estimation of an ordinal polythomous multidimensional LC IRT model by using data concerning a sample of 201 oncological
Italian patients who were asked to fill in questionnaires about their health and
perceived quality of life. Here we are interested in anxiety and
depression, as assessed by the ``Hospital Anxiety and Depression Scale'' (HADS)
developed by \cite{zig:sna:1983}.
The questionnaire is composed by $14$ polythomous items equally divided between the two
dimensions:

\begin{enumerate}
\item anxiety (7 items: 2, 6, 7, 8, 10, 11, 12);
\item depression (7 items: 1, 3, 4, 5, 9, 13, 14).
\end{enumerate}

All items of the HADS questionnaire have four response categories: the minimum value 0
corresponds to
a low level of anxiety or depression, whereas the maximum value 3 corresponds to a
high level of anxiety or depression.

Similarly to the  previously described example, we begin with loading data and aggregating them:
\begin{output}
> data(hads)
> X = as.matrix(hads)
> out1 = aggr_data(X) 
> S = out1$data_dis
> yv = out1$freq
\end{output}
Then, we define a suitable matrix to accommodate assumption of bidimensionality (we remind that unidimensionality structure of items is treated as default in the functions of \verb'MultiLCIRT' package):
\begin{output}
> dim2 = rbind(c(2,6,7,8,10,11,12),c(1,3,4,5,9,13,14)) 
\end{output}

To proceed to the selection of the optimal model (see Section \ref{sec:modsel}), we detect the optimal number $\hat{k}$ of latent classes. To this aim, we suggest to adopt the standard LC model \citep{good:74}. In this way, no choice on the link function and the item parameterization is requested; also, any restrictive assumptions on item dimensionality is avoided. Therefore, function \verb'est_multi_poly' is employed and a comparison among models which differ by the number of latent classes is performed for increasing values of $k$. As outlined in Section \ref{sec:model}, we rely on BIC index, taking as optimal number of latent classes that value just before the first increasing of BIC. 

\begin{output}
> out1 = est_multi_poly(S,yv,k=1,start=0, link=0)
> out2 = est_multi_poly(S,yv,k=2,start=0, link=0)
> out3 = est_multi_poly(S,yv,k=3,start=0, link=0)
> out4 = est_multi_poly(S,yv,k=4,start=0, link=0)
\end{output}

Results are obtained by calling outputs in objects \verb'out1' to \verb'out4' as follows, where maximum log-likelihood, number of model parameters, and BIC values for $k=1,\ldots,4$ latent classes are shown in columns one, two and three, respectively.

\begin{output}
> rbind(cbind(out1$lk, out1$np, out1$bic), cbind(out2$lk, out2$np, out2$bic), 
+ cbind(out3$lk, out3$np, out3$bic), cbind(out4$lk, out4$np, out4$bic)) 

          [,1]  [,2]  [,3]
[1,] -3153.151   42 6529.040
[2,] -2814.635   85 6080.051
[3,] -2677.822  128 6034.468
[4,] -2645.435  171 6197.736
\end{output}

On the basis of the adopted selection criterium, 
we choose $\hat{k} = 3$ as optimal number of 
latent classes as, in correspondence of this number of latent classes, the smallest 
estimated BIC value is observed. To avoid that the choice of $k$ falls in correspondence of a local - rather than a global - maximum point, we suggest to repeat the selection process by randomly varying the starting values  of the model parameters: for each possible value of $k$ the highest obtained log-likelihood value is identified and, consequently, the smallest estimated BIC value. For this aim, just put  \verb'start=1' rather than \verb'start=0' in function \verb'est_multi_poly'.

As regards to the following step concerning the choice of the best logit link function,  a 
comparison between a model with global logit link and a model with local logit link  is 
carried out on the basis of the BIC index and by assuming $\hat{k} = 3$ latent classes, free item discriminating and 
difficulties parameters, and a completely general multidimensional structure for the data 
(i.e., $r$ dimensions, one for each item).

\begin{output}
> out31 = est_multi_poly(S,yv,k=3,start=0, link=1,disc=1,difl=0, 
+ multi=cbind(1:ncol(S))) # Global logit
> out32 = est_multi_poly(S,yv,k=3,start=0, link=2,disc=1,difl=0, 
+ multi=cbind(1:ncol(S))) # Local logit
> cbind(rbind(out31$lk, out31$np, out31$bic), rbind(out32$lk, out32$np, 
+ out32$bic))
          [,1]      [,2]
[1,] -2726.348 -2741.321
[2,]    72.000    72.000
[3,]  5834.534  5864.479
\end{output}

Results show that a global logit link (first column)  has to be preferred to a local logit 
link (second column). Also, it can be observed that a graded response type model has a better fit than 
the standard LC model, as the BIC value observed for the former is smaller than that 
detected for the latter (6197.736).

Once we have chosen the global logit as the best link function, we carry on with the 
test of unidimensionality. 
An LR test is used to compare models which differ on account of their dimensional 
structure, all other elements being equal (i.e., free item discriminating and difficulty 
parameters), that is (\textit{i}) a graded response model with $r$-dimensional 
structure, (\textit{ii}) a graded response model with bidimensional structure (i.e., 
anxiety and depression), as suggested by the structure of the questionnaire, and (\textit{iii}) a graded response model with unidimensional 
structure (i.e., all the items belonging to the same dimension). The LR tests are performed through function \verb'test_dim', firstly applied to compare models at points (\textit{i}) and (\textit{ii}), and then applied to compare models at points (\textit{ii}) and (\textit{iii}).

\begin{output}
> test = test_dim(S, yv, k=3, link=1, disc=1, difl=0, multi0=dim2,
+ multi1=cbind(1:ncol(S)))

Log-likelihood of the constrained model =   -2731.7092 
Log-likelihood of the unconstrained model = -2726.3450
Deviance =                                   10.7226
Degrees of freedom =                         12 
P-value =                                   0.5528 

> test = test_dim(S, yv, k=3, link=1, disc=1, difl=0, multi1=dim2)

Log-likelihood of the constrained model =   -2731.8937
Log-likelihood of the unconstrained model = -2731.7092 
Deviance =                                   0.3689
Degrees of freedom =                         1 
P-value =                                   0.5436
\end{output}

The hypothesis of unidimensionality cannot be rejected. This
result is coherent with a similar analysis performed on the same data by \citep{bac:bart},
where item responses were dichotomized and a Rasch parameterization was adopted. We note that the test of unidimensionality may be substituted by a hierarchical clustering of items,  in case we should not have any idea of the grouping of items and of the latent traits measured by the questionnaire. 

As previously outlined, the choice of the number of parameters per item depends on both
the presence of a constant/non-constant discriminating index ($\gamma_j$), and of a
constant/non-constant threshold difficulty parameter ($\beta_{jx}$), for each item. In
our application, this implicates a comparison among four graded response-type models:  (\textit{i}) GRM properly called, (\textit{ii}) GRM with a rating scale parameterization (i.e., free discriminating item parameters and constrained difficulty item parameters) or RS-GRM \citep{mur:90}, (\textit{iii}) GRM with constrained discriminating parameters and free difficulty parameters, also known as 1P-GRM \citep{ark:01}, (\textit{iv}) GRM with both constrained discriminating and difficulty parameters or 1P-RS-GRM \citep{ark:01}.
The parameterization is chosen on account of the unidimensional data structure and the
previously selected global logit link function. Besides, because the compared models are
nested, the parameterization is selected on the basis of an LR test. For the sake 
of completeness, log-likelihood and BIC
values are also provided for each model considered.

\begin{output}
# Unidimensional GRM
> out311 = est_multi_poly(S, yv, k=3, start=0, link=1, disc=1, difl=0)  
# Unidimensional  RS-GRM 
> out3111 = est_multi_poly(S, yv, k=3, start=0, link=1, disc=1, difl=1)   
# Unidimensional  1P-GRM
> out3112 = est_multi_poly(S, yv, k=3, start=0, link=1, disc=0, difl=0)   
# Unidimensional  1P-RS-GRM 
> out3113 = est_multi_poly(S, yv, k=3,start=0, link=1, disc=0, difl=1)   

# comparison between RS-GRM vs GRM
> Dev1 = -2*(out3111$lk - out311$lk)  
> df1 = out311$np - out3111$np
> Pv1 = 1-pchisq(q = Dev1, df1)

# comparison between 1P-GRM vs GRM
> Dev2 = -2*(out3112$lk - out311$lk)  
> df2 = out311$np - out3112$np
> Pv2 = 1-pchisq(q = Dev2, df2)

# comparison between 1P-RS-GRM vs 1P-GRM
> Dev3 = -2*(out3113$lk - out3112$lk)  
> df3 = out3112$np - out3113$np
> Pv3 = 1-pchisq(q = Dev3, df3)

> rbind(cbind(disc=1, difl=0, lk=out311$lk, np=out311$np, BIC=out311$bic,  
+ LR="", p="", " "),
+ cbind(disc= 1, difl=1, out3111$lk, out3111$np, out3111$bic, Dev1, Pv1,
+ "(RS-GRM vs GRM)"),
+ cbind(disc=0, difl=0, out3112$lk, out3112$np, out3112$bic, Dev2, Pv2,
+ "(1P-GRM vs GRM)"),
+ cbind(disc=0, difl=1, out3113$lk, out3113$np, out3113$bic, Dev3, Pv3,
+ "(1P-RS-GRM vs 1P-GRM)"))

     disc difl    lk     np   BIC       LR     p   
[1,]    1    0 -2731.894 59 5776.682   NaN    NaN  " "      
[2,]    1    1 -2795.570 33 5766.149 127.353 0.000 "(RS-GRM vs GRM)"
[3,]    0    0 -2741.285 46 5726.521  18.782 0.130 "(1P-GRM vs GRM)"
[4,]    0    1 -2844.518 20 5795.102 206.467 0.000 "(1P-RS-GRM vs 1P-GRM)"
\end{output}

The analyses show  that between GRM and RS-GRM, GRM has to be
preferred to RS-GRM, while between models GRM and 1P-GRM, the latter has to be
preferred. Besides, as model 1P-GRM has a better fit than model 1P-RS-GRM, then 1P-GRM
has to be preferred model among the four considered, that is the graded response type
models with free $\beta_{jx}$ parameters 
and constant $\gamma_j$ parameters. Such a result is achieved by
taking into account both the BIC criterium and the LR test.

As the sequence of the previously described steps may be considered partly arguable, it
can be also shown that the same results - in terms of link function, item
parameterization and dimensionality choice - would have been obtained if each of such
models were compared at once accounting for log-likelihood and BIC values as selection
criteria. Indeed, Table \ref{table9} shows that the smallest BIC value is observed when
selecting: (\textit{i}) a global logit link function; (\textit{ii}) constrained
$\gamma_j$ parameters and
free $\beta_{jx}$ parameters, that is, a 1P-GRM model; and (\textit{iii}) assuming a
unidimensional structure for the data.

\begin{table}[!ht]\centering
\vspace*{0.5cm}
{\small
\begin{tabular}{c|c|c|cc|cc}
\hline\hline
Dimensionality & \multicolumn{2}{c|}{Item parameters} &     \multicolumn{2}{c|}{Global
logit}      & \multicolumn2c{Local logit}   \\
\hline
           & \multicolumn1c{$\gamma_j$}
           & \multicolumn{1}{c|}{$\beta_{jx}$}
           &       \multicolumn1c{$\hat{\ell}$}   &
           \multicolumn{1}{c|}{BIC}        &
           \multicolumn1c{$\hat{\ell}$}    &
           \multicolumn1c{BIC}     \\
\hline
$r$-dimensional 	&	free/constr.	&	free	&	-2726.347   &  5834.534 & -2741.321   
& 5864.479  	\\	            
	                        &	free/constr.	&	constrained	& -2815.568	&  5875.088 
	                        & -2836.766   &  5917.484  \\
	                        \hline
bidimensional 	&	free	&	free	&	-2731,249	&	5780,696	&	-2749,839	
&	 5817,877	\\
	&	constrained	&	free	&	-2740,658	&	5735,875	&	-2764,787	&	
	5784,132	 \\
	&	free	&	constrained	&	-2798,959	&	5778,230	&	-2835,611	&	
	5851,534	 \\
	&	constrained	&	constrained	&	-2843,227	&	5803,127	&	-2869,223	&	
	5855,120	\\
\hline
unidimensional 	&	free	&	free	&	-2731,894	&	5776,682	&	-2750,214
	&	 5813,323	\\
	&	constrained	&	free	&	-2741,285	&	\bf 5726,521	&	-2765,129	&	
	5774,211	\\
	&	free	&	constrained	&	-2795,570	&	5766,149	&	-2833,179	&	
	5841,366	 \\
	&	constrained	&	constrained	&	-2844,518	&	5795,102	&	-2870,178	&	
	5846,422	\\
\hline
\end{tabular}}
\caption{\em Log-likelihood and BIC values for the global and local logit link
functions, taking into account the dimensional structure ($r$-dimensional/bidimensional/
unidimensional)
and the item parameters (depending on whether they are free/constraint); in boldface is
the smallest BIC value.}
\label{table9}\vspace*{0.5cm}
\end{table}

The estimates of support points $\hat{\xi}_c$ and probabilities $\hat{\pi}_c$,
$c=1,2,3$, under the selected unidimensional 1P-GRM model are extrapolated from object \verb'out3112'.  

\begin{output}
> rbind(out3112$Th, out3112$piv)
           [,1]      [,2]      [,3]
[1,] -0.7757623 1.1830799 3.4185177
[2,]  0.3419038 0.4911631 0.1669331
\end{output}

On the basis of these results, we conclude
that patients who suffer from
psychopatological disturbs are mostly represented in the first two classes, whereas only
the $16.7\%$ of the subjects belong to the third class. Furthermore, patients belonging
to class 1 present the least severe conditions, whereas
patients in class 3 present the worst conditions. 

To conclude, we outline that the proposed model selection procedure results to be simplified in case of binary items, being the selection of logit link function and the selection of item difficulties not requested.

\section{Conclusions}\label{sec:conclusion}
In this paper, we initially illustrate a class of models that extends traditional IRT models allowing for ({\em i}) multidimensionality, ({\em ii})  discreteness of latent traits, and  ({\em iii}) item responses of different nature, in a same unifying framework. 
In particular, the assumption of multidimensionality allows
us to take more than one latent trait into account at the same time and to study the
correlation between latent traits. Moreover, through the introduction of latent classes,  homogeneous
subpopulations of subjects are detected and  a notable simplification from the computational point of view results with respect to the case of continuous latent traits, where the marginal likelihood is characterized by a multidimensional integral difficult to treat.  Lastly, the proposed class of models is formulated to treat with both binary and ordinal polythomous items and, in the latter case, different link functions and item parameterizations are provided.

In order to make inference on the proposed model, we show how the
log-likelihood may be efficiently maximized by the EM algorithm. We 
also propose a model selection procedure to choose the
different features that contribute to define a specific multidimensional LC IRT model.
In general, comparisons between different parameterizations are based on information
criteria, in particular we rely on the Bayesian Information Criterion, 
or on likelihood ratio test, being this last tool
useful in presence of nested models.

The estimation of the multidimensional LC IRT models is made possible by the availability of \textsf{R} package \verb'MultiLCIRT' \citep{multi:pack}, the use of which is illustrated through the description of the main functions and two applications to real data. The first application is about data collected by the Educational Testing Service within the National Assessment of Educational Progress (NAEP) project and it implements the model-based hierarchical  clustering procedure for a set of binary items. The results are coherent with those obtained by \cite{bart:07}, showing that items may be grouped in three homogeneous groups. The second application is about  data concerning the measurement of psychopathological disturbs through the Hospital Anxiety and Depression Scale (HADS). By this application we show how performing the model selection in presence of ordinal polythomous data. The results show that subjects can be classified in three latent classes  and the item responses can
be explained by a graded response type model with items having the same discriminating
power and different distances between consecutive response categories. The
bidimensionality assumption is rejected in favor of unidimensionality, so that all
items of the questionnaire measure the same latent  psychopathological disturb. All \textsf{R} replication scripts for the proposed examples are provided. 

\bibliography{biblio}
\bibliographystyle{apalike}

\end{document}